\documentclass[%
reprint,
 amsmath,amssymb,
 aps,
prb,
]{revtex4-2}

\usepackage{graphicx}
\usepackage{dcolumn}
\usepackage{bm}

\usepackage{comment}
\usepackage{color}

\begin{document}

\preprint{APS/123-QED}

\title{Kondo Temperature Evaluated from Linear Conductance in Magnetic Fields}

\author{Rui Sakano}
\email{sakano@keio.jp}
\affiliation{%
Faculty of Science and Technology, Keio University, 3-14-1 Hiyoshi, Kohoku-ku, Yokohama 223-8522, Japan
}

\author{Tokuro Hata}
\affiliation{
Department of Physics, Tokyo Institute of Technology, 2-12-1 Ookayama, Meguro, Tokyo 152-8551, Japan.
}

\author{Kaiji Motoyama}
\affiliation{
Department of Physics, Osaka City University, Sumiyoshi-ku, Osaka, 558-8585, Japan
}

\author{Yoshimichi Teratani}
\affiliation{
Department of Physics, Osaka City University, Sumiyoshi-ku, Osaka, 558-8585, Japan
}
\affiliation{
Nambu Yoichiro Institute of Theoretical and Experimental Physics,
Osaka Metropolitan University, Sumiyoshi-ku, Osaka, 558-8585, Japan
}

\author{Kazuhiko Tsutsumi}
\affiliation{
Department of Physics, Osaka City University, Sumiyoshi-ku, Osaka, 558-8585, Japan
}
\affiliation{
Nambu Yoichiro Institute of Theoretical and Experimental Physics,
Osaka Metropolitan University,
Sumiyoshi-ku, Osaka, 558-8585, Japan
}

\author{Tomonori Arakawa}
\affiliation{
National Institute of Advanced Industrial Science and Technology, National Metrology Institute of Japan, Tsukuba, Ibaraki 305-8563, Japan
}
\author{Meydi Ferrier}
\affiliation{Universit\'{e} Paris-Saclay, CNRS, Laboratoire de Physique des Solides, 91405, Orsay, France}

\author{Richard Deblock}
\affiliation{Universit\'{e} Paris-Saclay, CNRS, Laboratoire de Physique des Solides, 91405, Orsay, France}

\author{Mikio Eto}
\affiliation{
Faculty of Science and Technology, Keio University, 3-14-1 Hiyoshi, Kohoku-ku, Yokohama 223-8522, Japan
}

\author{Kensuke Kobayashi}
\affiliation{
Institute for Physics of Intelligence and Department of Physics,
The University of Tokyo, Bunkyo-ku, Tokyo 113-0033, Japan.}

\author{Akira Oguri}
\affiliation{
Department of Physics, Osaka City University, Sumiyoshi-ku, Osaka, 558-8585, Japan
}
\affiliation{
Nambu Yoichiro Institute of Theoretical and Experimental Physics,
Osaka Metropolitan University,
Sumiyoshi-ku, Osaka, 558-8585, Japan
}%



\date{\today}

\begin{abstract}
We theoretically and experimentally study
the universal scaling property of the spin-$\frac{1}{2}$ Kondo state
in the magnetic field dependence of
bias-voltage linear conductance
through a quantum dot
at low temperatures.
We discuss
an efficient and reliable
procedure to evaluate
the Kondo temperature defined at the ground state
from experimental or numerical data sets of
the magnetic field dependence of the linear conductance or the magnetization of the quantum dot.
This procedure is
helpful for quantitative comparison of the theory and the experiment,
and useful in Kondo-correlated systems
where temperature control over a wide range is difficult,
such as for cold atoms.
We demonstrate its application to experimentally
measure
electric current
through a carbon nanotube quantum dot.
\end{abstract}

\maketitle

\section{introduction}

The Kondo effect is a typical many-body phenomenon of electrons in metals
and have been thoroughly studied for over 60 years
\cite{hewson_1993, kondo_2012}.
It is a phenomenon in which
a localized magnetic moment of a magnetic impurity, or quantum dots, is screened by conduction
electrons to form a many-body singlet ground state.
Simultaneously, 
the Kondo resonance state emerges 
in the impurity or the quantum dot at the Fermi level of the conduction electrons.
In quantum dots, the tunneling through the resonance state enhances the electric current
between two leads 
at low temperatures.
The Kondo effect in quantum dots has been experimentally observed
as an increase of bias-voltage linear conductance,
with lowering the temperature in dilution refrigerators
\cite{Nature.391.156,PhysRevLett.81.5225,doi:10.1126/science.289.5487.2105}.
Recently, the Kondo effect
has been intensively investigated in various physical systems,
such as dilute magnetic alloys, heavy fermions
\cite{hewson_1993},
quantum dots
\cite{Nature.391.156,PhysRevLett.81.5225,doi:10.1126/science.289.5487.2105},
cold atoms
\cite{PhysRevLett.111.135301,PhysRevA.99.032707,
PhysRevLett.120.143601,PhysRevLett.123.143002},
and quark matter
\cite{PhysRevD.96.114007,PhysRevD.96.114027}.

A prominent feature of the Kondo effect is the scaling universality
in the temperature dependence of physical quantities and their response to external fields
at lower energies than a scaling energy.
The scaling energy is called the Kondo temperature.
At much lower temperatures that the Kondo temperature,
the local Fermi liquid describes the properties of the Kondo effect,
which is an extension of Landau's Fermi liquid to impurity or quantum dot systems
\cite{Nozieres1974,J.Phys.Cond.Matt.13.10011}.
In the local Fermi liquid theory,
only one parameter given by the Kondo temperature describes the low energy state.
Namely, the Kondo temperature is the essential quantity
to understand the low energy properties of the Kondo effect.

The Kondo temperature can be deduced from low-energy properties, 
such as the ground-state value of magnetic susceptibility,
the coefficient of the $T$-linear specific heat,
or the width of the Kondo resonance.
In experiments, the Kondo temperature is usually determined
by the energy scale of the crossover transition 
between low and high energies.
However, it is challenging to specify the crossover transition
point, such that it is quantitatively consistent
with the theoretically defined Kondo temperature.
Evaluation of the Kondo temperature
that is quantitatively consistent with theory
is the key to exploring the frontier of the local Fermi liquid nature
of the many-body state,
such as nonlinear current and current due to the Kondo correlation
\cite{PhysRevLett.100.246601,PhysRevB.79.165413,PhysRevB.84.245316,PhysRevB.85.201301,Nat.Phys.12.230,PhysRevLett.118.196803,Nat.Comm.12.3233,PhysRevLett.128.147701}. 
It is also demanded
to investigate systems in which the Kondo effect competes
with other correlation effects,
such as superconductivity and spin-orbit interaction
\cite{PhysRevLett.99.136806,PhysRevLett.104.076805,PhysRevB.81.121308,
PhysRevLett.121.247703,Kanai2011}.
This paper aims to show
an efficient and reliable procedure
to evaluate the Kondo temperature that is quantitatively consistent
with local Fermi liquid theory
from a data set of magnetic field dependence of linear conductance
through a quantum dot.
Our procedure is extendable to magnetic field dependence of other quantities
in a variety of Kondo correlated systems.

In 1998,
Goldhabar-Gordon {\it et al.} devised an empirical formula for
the temperature dependence of the linear conductance
of electric current
through a quantum dot connecting to two electric leads
\cite{Nature.391.156,PhysRevLett.81.5225,PhysRevB.84.245316,PhysRevB.85.201301}.
They assume a simple analytic function
\begin{align}
    G_{\rm emp}(T) =
    G_0^{} \left[
    1+\left( \frac{T}{\widetilde{T}_{\rm K}^{\rm emp}} \right)^2
    \right]^{-s}
    \label{eqn:empT}
\end{align}
as the temperature dependence of the linear conductance.
Here, $G_0^{}=\frac{2e^2}{h} \frac{4\Gamma_L^{}\Gamma_R^{}}{(\Gamma_L^{}+\Gamma_R^{})^2}$,
and the Kondo temperature is empirically determined by
$\widetilde{T}_{\rm K}^{\rm emp}=T_{\rm K}^{\rm emp}/\sqrt{2^{1/s}-1}$
so that
$G_{\rm emp}^{} (T_{\rm K}^{\rm emp}) = G_{0}^{} /2$.
$\Gamma_{L}^{}$($\Gamma_{R}^{}$) is the linewidth due to the electron tunneling
between the left (right) lead and the dot,
as defined in the next section.
The curve fitting to a set of data calculated
by the numerical renormalization group (NRG) approach
for the spin-$\frac{1}{2}$ Anderson impurity model determines
the fitting parameter as $s=0.22$.
Mesoscopic physicists have frequently applied the curve fitting
to experimental data
to evaluate the Kondo temperature,
taking Eq.~\eqref{eqn:empT} as the model function
and $T_{\rm K}^{\rm emp}$ as the fitting parameter.
Kretinin {\it et al.} showed that this empirical Kondo temperature relates to the Kondo temperature
defined in the local Fermi liquid theory
$T_{\rm K}^{}$ given in Eq.~\eqref{eq:TK1}
as
\begin{align}
    T_{\rm K}^{\rm emp} = \eta(U) \, T_{\rm K}^{}
    \label{eqn:modeldependence}
\end{align}
with
$\eta (U)$
that depends on the Coulomb interaction $U$ at the dot site
\cite{PhysRevB.84.245316}.
Namely, the empirically evaluated Kondo temperature demands
an extra factor
to be consistent with the Kondo temperature
defined in the local Fermi liquid theory.

In this paper,
we theoretically investigate the scaling universality
in magnetic field dependence of the linear conductance of a quantum dot,
using the numerical renormalization group approach
and the Bethe ansatz exact solution.
Strong magnetic fields suppress the Kondo effect,
as it has also been observed in transport experiments of quantum dots
\cite{Nature.391.156,PhysRevLett.81.5225,doi:10.1126/science.289.5487.2105,PhysRevB.74.233301,Bauer2013,doi:10.1021/acs.jpclett.1c01544}.
Exploiting the universal curve of the magnetic-field scaling at $T=0$,
we develop
a simple procedure to evaluate the Kondo temperature
of the local Fermi liquid theory
from magnetic field dependence of the linear conductance,
while Goldhabar-Gordon's approach given by Eq.~\eqref{eqn:empT}
uses the excited state.
Then, we examine
our procedure
for experimental data on electric current
measured in a carbon nanotube quantum dot.

There have already been a lot of preceding works
on magnetotransport and magnetic properties
in the Kondo impurity and the Kondo dot
\cite{doi:10.1143/JPSJ.51.3192,doi:10.1143/JPSJ.52.1119,PhysRevB.64.241310,PhysRevB.66.125304,Pustilnik_2004,PhysRevB.74.233304,PhysRevB.87.165132,Bauer2013,PhysRevB.98.075404,PhysRevLett.120.126802,PhysRevB.97.035435,PhysRevLett.125.216801,doi:10.1021/acs.jpclett.1c01544,PhysRevB.105.115409}.
We here develop a way to evaluate the Kondo temperature
with the use of
the universal curve of
the {\it bias-voltage-linear} magnetoconductance.
In our previous work,
we evaluated the Kondo temperature
using the universal curve of the bias-voltage
{\it bias-voltage-nonlinear} magnetoconductance
\cite{Nat.Comm.12.3233}.
This curve fitting of the linear conductance still has
the following advantages.
The linear conductance is a monotonous function of the applied magnetic field,
as seen in this paper,
while the nonlinear one is nonmonotonous.
Naively, the measurement and evaluation of the Kondo temperature
for the linear conductance is
simpler and easier than that for the nonlinear one in experiments.
In particular,
the Kondo temperature evaluated from the magnetic field dependence
in our approach
is quantitatively consistent with the one defined by the linewidth of the quasiparticle in the local Fermi liquid theory.
In semiconductor nano-devices experiments,
manipulation of the magnetic field is more stable and easy than temperature control.
Indeed, it is technically demanding to systematically address the temperature between 1K and 4K
in dilution refrigerators.
In addition, changing temperature could be more challenging
in other systems, such as cold atoms.
Therefore, our method using magnetic fields
is promising for extending the research field of the Kondo physics.

This paper is organized as follows.
In Sec.~\ref{sec:ModelAndCalculation},
we introduce our model of a quantum dot system and explain the calculation of the linear conductance.
In Sec.~\ref{sec:ResultAndDiscusstion},
we discuss the universal behavior of the Kondo state
on the magnetic field dependence of the linear conductance
through a quantum dot 
and our way to estimate the Kondo temperature from the magnetic field dependence.
Then, in Sec.~\ref{sec:ResultAndDiscusstion}, we demonstrate its application to experimental data
measured in a carbon nanotube quantum dot.
Finally, in Sec.~\ref{sec:summary}, we summarize the paper.

\section{model and calculation}\label{sec:ModelAndCalculation}

We introduce the model to describe a quantum dot,
where the Kondo state enhances the electric current.
We also define the Kondo temperature in the local Fermi liquid theory.
Then we show an analytical result of the linear conductance
under applied magnetic fields,
using the Bethe ansatz exact solution.

\subsection{Model}\label{subsec:Model}
Let us consider a quantum dot connected to two electric leads,
which is described by
the spin-$\frac{1}{2}$ Anderson impurity model
\begin{align}
    \cal{ H}_{\rm A}^{}=\cal{ H}_{\rm d}^{} + \cal{ H}_{\rm c}^{} + \cal{ H}_{\rm T}^{}
    \label{eqn:AIM}
\end{align}
with
\begin{align}
	\cal{ H}_{\rm d}^{} &=
	\sum_{\sigma} \epsilon_{{\rm d}} n_{{\rm d}\sigma} ^{}
	- \mu_{\rm B}B \, m_{\rm d}^{}
		+U n_{{\rm d}\uparrow}^{} n_{{\rm d}\downarrow}^{}
	\, ,\\
	\cal{ H}_{\rm c}^{} &=
	\sum_{\gamma\sigma} \int_{-D}^D d\varepsilon \,\varepsilon c_{\varepsilon \gamma \sigma}^{\dagger} c_{\varepsilon \gamma \sigma}^{}
	\, , \\
	\cal{ H}_{\rm T}^{} &=
	\sum_{\gamma \sigma} \left(
	v_{\gamma}^{} d_{\sigma}^{\dagger} \psi_{\gamma\sigma}
	+ v_{\gamma}^* \psi_{\gamma \sigma}^{\dagger} d_{\sigma}^{}
	\right)
	\, .
\end{align}
${\cal H}_{\rm d}$ represents that
electrons interact with each other by the Coulomb interaction $U$
at the energy level of the quantum dot
$\epsilon_{\rm d}^{}$
under the applied magnetic field $B$.
$n_{{\rm d}\sigma}^{}(:=d_{\sigma}^{\dagger}d_{\sigma}^{})$
is the operator of the electron occupation at the dot site
and
$m_{\rm d}(:=n_{{\rm d}\uparrow}^{} - n_{{\rm d}\downarrow}^{})$
is the magnetization,
where the operator $d_{\sigma}$ annihilates an electron with energy $\epsilon_{\rm d}^{}$ and spin $\sigma=\uparrow, \downarrow$
at the dot-site.
$\mu_{\rm B}^{}$ is the Bohr magneton and
we set the $g$-factor as
$g=2$.
${\cal H}_{\rm c}^{}$
represents the conduction electrons in the left lead $\gamma=L$ and the right lead $\gamma=R$
with half width of the conduction band $D$.
The operator $c_{\varepsilon\gamma\sigma}^{}$ annihilates an electron
with energy $\varepsilon$ and spin $\sigma$
in the lead $\gamma$,
and satisfies with the anti-commutation relation
$\left\{c_{\varepsilon \gamma \sigma}^{\dagger}, c_{\varepsilon'\gamma' \sigma'}^{} \right\}
= \delta(\varepsilon -\varepsilon') \, \delta_{\gamma\gamma'} \, \delta_{\sigma\sigma'}^{}$.
${\cal H}_{\rm T}^{}$ represents
electron tunneling between the dot and the leads
via the tunneling matrix $v_{\gamma}^{}$ and
$\psi_{\gamma\sigma}^{} := \sqrt{\rho_{\rm c}^{} } \int_{-D}^D d\varepsilon \,c_{\varepsilon\gamma\sigma}^{}$
with $\rho_{\rm c}^{}$ the density of states of the conduction electrons.
The linewidth of the dot level due to the electron tunneling is
\begin{align}
\Delta := \frac{1}{2} \left( \Gamma_L^{} + \Gamma_R^{} \right), \quad
\Gamma_{\gamma}^{} := 2 \pi \rho_{\rm c}^{} v_{\gamma}^{2}
\,.
\end{align}
We measure $\epsilon_{\rm d}^{}$ and $\varepsilon$ from the chemical potential of the leads at equilibrium
$\mu_L^{}=\mu_R^{}=0$.  
We use
$\hbar=k_{\rm B}^{}=1$
throughout this paper.

In this paper,
we define the Kondo temperature
by $\widetilde{\Delta}$
the renormalized linewidth of the quasiparticle of the local Fermi liquid
in the form
\begin{align}
    T_{\rm K}^{} := \frac{\pi\widetilde{\Delta}}{4}.
    \label{eq:TK1}
\end{align}
The linewidth of the quasiparticle is given by the wave function renormalization factor $z$ as
$\widetilde{\Delta} = z \Delta$
in the microscopic theory.
The renormalization factor
\begin{align}
z= \left[ 1 -\left. \frac{\partial \Sigma_{{\rm d}\sigma}^{\rm r} (\omega)}{\partial \omega}\right|_{\omega=0} \right]^{-1}
\end{align}
is given by $\Sigma_{{\rm d}\sigma}^{\rm r} (\omega)$
the self-energy of the retarded
Green's function for the electrons in the dot
at absolute zero.
We note that
the Kondo temperature is also determined 
by the spin susceptibility
$\chi_{\rm s}^{} = \mu_{\rm B}^{} \partial \langle m_{\rm d}^{}\rangle/\partial B|_{B=0,T=0}^{}$
or the $T$-linear specific heat coefficient $\gamma_{\rm d}^{} = \lim_{T \to 0} C(T)/T$.
Here, $m_{\rm d}^{} = n_{{\rm d}\uparrow}^{} -n_{{\rm d}\downarrow}^{}$ is the magnetization at the dot,
and $C(T)$ is the specific heat.
This definition is consistent with Eq.~\eqref{eq:TK1} in the Kondo regime:
$ \chi_{\rm s}^{} / { \mu_{\rm B}^{} }^2 = (6/\pi^2) \gamma_{\rm d}^{} = T_{\rm K}^{-1}$ in the Kondo limit.

\subsection{Linear conductance}
The linear conductance of the electric current through the quantum dot
with applied bias voltages between leads $L$ and $R$
at absolute zero in the magnetic fields is given
in the form
\begin{align}
    G = \frac{G_0^{}}{2} \sum_{\sigma}
    \sin^2 \delta_{\sigma}^{}
    \, .
\end{align}
The conductance is given only by the phase shift $\delta_{\sigma}^{}$
that describes the Kondo ground state.
Friedel's sum rule relates the phase shift $\delta_{\sigma}^{}$
to the number of the electrons at the dot as
$\delta_{\sigma}^{}=\pi\left\langle n_{{\rm d}\sigma}^{}\right\rangle$.
At the half-filling point
$\epsilon_{\rm d}^{} = - U/2$,
the conductance is given 
by the magnetization at the dot
$\langle m_{\rm d}^{}\rangle$ in a simple form
\cite{PhysRevB.98.075404}
\begin{align}
    G = G_0^{} \cos^2 \left( \frac{\pi}{2} \langle m_{\rm d}^{}\rangle \right)
    \label{eqn:ConductanceInMFAtPHS}
\end{align}
because the particle-hole symmetry fixes the total number of the electrons
at the dot site as
$\langle n_{\rm d}^{}\rangle =1$
with $n_{\rm d}^{} = n_{{\rm d}\uparrow}^{} + n_{{\rm d}\downarrow}^{}$.
We use the numerical renormalization group approach
and the Bethe ansatz exact solution (BAES) to calculate the electron occupation and  the magnetization
at the dot.

Next, we consider extending the applicability of
Eq.~\eqref{eqn:ConductanceInMFAtPHS} to the Kondo regime
where
$\Delta \ll - \epsilon_{\rm d}^{}$, and $\Delta \ll U + \epsilon_{\rm d}^{}$.
The first inequality means that either one or two electrons occupy the dot,
while the second inequality excludes two-electrons occupation.
The charge fluctuation is drastically suppressed as $\chi_{\rm c}^{}/\chi_{\rm s}^{} \ll 1$,
with $\chi_{\rm c}^{}= - \frac{\partial \langle n_{\rm d}^{} \rangle}{\partial\epsilon_{\rm d}^{}}$ the charge susceptibility.
This results in the total number of electrons at the dot site being locked
at $\langle n_{\rm d}^{} \rangle \simeq 1.0$ in the Kondo regime.
Therefore, the conductance in the Kondo regime is given in the form
$G \simeq G_0^{} \cos^2 \left( \frac{\pi}{2} \langle m_{\rm d}^{}\rangle \right)$
with Eq.~\eqref{eqn:ConductanceInMFAtPHS}.

In the Kondo regime
we have the exact solution for the Kondo model
 (see Appendix \ref{Appendix:KondoModel})
\cite{RevModPhys.55.331}.
The magnetic field dependence of the magnetization is given
in simple series and integral
as
\begin{align}
	\langle m_{\rm d}^{} \rangle =
	\sum_{n=0}^{\infty} \frac{(-1)^n}{n!} \left(\frac{\pi}{4}\right)^n (2n+1)^{n-\frac{1}{2}} \left( \frac{ \mu_{\rm B}^{} B }{T_{\rm K}^{}} \right)^{2n+1}
	\label{eqn:md4smallB}
\end{align}
for $\mu_{\rm B} B \le \sqrt{2/(e\pi)} \, T_{\rm K}^{}$,
and
\begin{align}
	& \langle m_{\rm d}^{}\rangle
	\nonumber \\
	& =
	1 - \pi^{-\frac{3}{2}} \int_0^{\infty} dx \, \sin(\pi x) \frac{\Gamma(x+\frac{1}{2})}{x^{x+1}} \left( \frac{2}{\pi}\right)^x \left( \frac{\mu_{\rm B} B}{T_{\rm K}^{}} \right)_{}^{-2x} 
	\label{eqn:md4largeB}
\end{align}
for $\mu_{\rm B} B > \sqrt{2/(e\pi)} \, T_{\rm K}^{}$.
Here, we note that $\sqrt{2/(e\pi)} = 0.4839\cdots$.

Using this exact result,
we also derive
the asymptotic form of the conductance for small and large magnetic fields.
For small magnetic fields
$\mu_{\rm B}^{} B \ll T_{\rm K}^{}$,
the linear conductance
is given
up to the leading $B$-dependent term
in the form
\cite{PhysRevB.66.125304,PhysRevB.87.165132}
\begin{align}
    G (B) & = G_0^{} \left[
    1 - \frac{\pi^2}{4} \left( \frac{\chi_{\rm s}^{}}{\mu_{\rm B}^{}} B \right)^2  
    \right]
    + {\cal O}\left( \left( \frac{\chi_{\rm s}^{}}{\mu_{\rm B}^{}} B \right)^4 \right)
    \nonumber \\
    & = G_0^{} \left[
    1 - \frac{\pi^2}{4} \left( \frac{\mu_{\rm B}^{} B }{ T_{\rm K}^{} } \right)^2
    \right]
    + {\cal O}
    \left( \left( \frac{\mu_{\rm B} B }{T_{\rm K} } \right)^4  \right)
    \, . \label{eqn:asympototicG4LF}
\end{align}
For large fields $\mu_{\rm B}^{} B \gg T_{\rm K}^{}$,
the conductance up to the leading $B$ dependence is given in the form
\begin{align}
    G(B) \sim G_0^{} \frac{\pi^2}{16}
    \left[
    \ln \left(\frac{ \mu_{\rm B}^{} B }{\sqrt{2/(e \pi )}T_{\rm K}^{}} \right)
    \right]^{-2}
    \, . \label{eqn:asympototicG4HF}
\end{align}

\section{result and discussion}\label{sec:ResultAndDiscusstion}

We theoretically discuss the universal scaling property of
the linear magnetoconductance.
Then, using the universal scaling property,
we devise a procedure to evaluate the Kondo temperature.
We finally demonstrate its application to experimental data
of conductance of a carbon nanotube quantum dot.

\subsection{Universal scaling of conductance under magnetic field}

\begin{figure*}[tb]
\includegraphics[scale =0.50
]{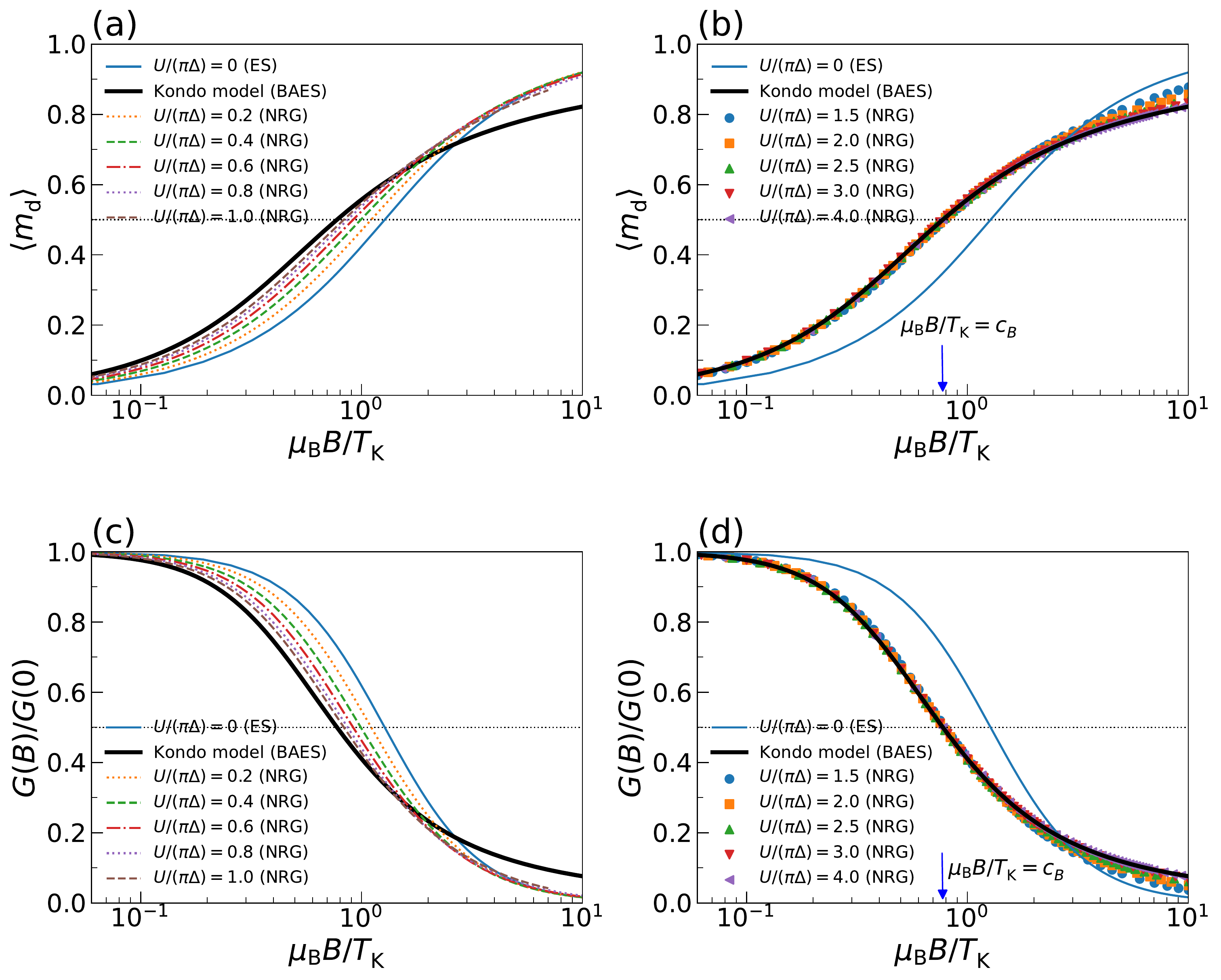}
\caption{\label{fig:G-B_log}
Magnetization $\langle m_{\rm d}^{} \rangle$
at particle-hole symmetric point
($\epsilon_{\rm d}^{} = - U/2$)
as a function of magnetic field $B$ at absolute zero for
(a) small $U/(\pi \Gamma) (\le 1)$ and
(b) large $U/(\pi \Gamma) (\le 1.5)$.
The resulting linear conductance $G$
for
(c) small $U/(\pi \Gamma) (\le 1)$ and
(d) large $U/(\pi \Gamma) (\le 1.5)$.
The broken lines and the data points
are calculated by the NRG approach. 
The thin (blue) solid lines are the conductance at absolute zero
for the noninteracting limit $U=0$,
which are calculated by the exact solution (ES).
The thick solid (black) lines are the conductance for the Kondo limit
($U \gg \Gamma$),
which are calculated by the Bethe ansatz exact solution.
In this figure,
we use the Kondo temperature $T_{\rm K}^{}$ of the local Fermi liquid theory,
given by Eq.~\eqref{eq:TK1} at $B=0$.
The arrow in (b) and (d) indicates the magnetic field
at which the conductance and the magnetization
takes half of the zero field value.
}
\end{figure*}

We show the magnetization and the linear conductance as a function of the applied magnetic field $B$
at the half-filling point
$\epsilon_{\rm d}^{} = - U/2$
and $T=0$
in Fig.~\ref{fig:G-B_log}.
We used the numerical renormalization group approach for finite
$U$'s,
the exact result for the noninteracting limit $U = 0$,
and the Bethe ansatz exact solution for the Kondo model $U \gg \Delta$
to calculate the quantities.
The magnetic field is scaled
by the Kondo temperature determined by Eq.~\eqref{eq:TK1}
at the zero field $B=0$ for each strength of interaction $U$,
to investigate the universal scaling properties of the Kondo effect. 

For a small Coulomb interaction $U/(\pi\Delta) \le 1$,
the curve of the magnetic field dependence of the magnetization and the conductance
continuously transits
between the noninteracting limit and the Kondo limit,
as seen in Figs.~\ref{fig:G-B_log}(a) and \ref{fig:G-B_log}(c).
The data points of the magnetic field dependence of the magnetization and the conductance
for a large Coulomb interaction $U/(\pi\Delta) \ge 1.5$
collapse to a single curve that is given by the Kondo model
for $\mu_{\rm B}^{} B< T_{\rm K}^{}$
as seen in Figs.~\ref{fig:G-B_log}(b) and \ref{fig:G-B_log}(d),
because of the Kondo universal scaling property.
However, for large magnetic fields $\mu_{\rm B}^{} B > T_{\rm K}^{}$,
the spin fluctuation becomes less pronounced,
and the data points consequently deviate from the curve of the Kondo model
to the curve for the noninteracting limit $U/(\pi\Delta)=0$.

Let us consider
a procedure to evaluate the Kondo temperature
from a measurement of the magnetic field dependence
of the linear conductance.
As seen in Figs.~\ref{fig:G-B_log}(b) and \ref{fig:G-B_log}(d),
the magnetic field $\bar{B}$
at which the conductance takes the half value of the zero-field value
$G(\bar{B} ) = G(0)/2$
stays on the universal $\mu_{\rm B}^{} B/T_{\rm K}^{}$ scaling curve.
Therefore, at this point,
the universal function gives the following relationship
between the magnetic field $\bar{B}$ and the Kondo temperature:
\begin{align}
    \frac{1}{2} g \mu_{\rm B}^{} \bar{B} = c_B^{} \, T_{\rm K}^{}
    \, .
    \label{eq:TKandBbar}
\end{align}
Here, we reinserted the $g$-factor.
We numerically calculate the coefficient $c_B^{}$
using the exact solution
for the Kondo model
given by Eq.~\eqref{eqn:md4largeB} with
three significant figures, as
\begin{align}
    c_B^{}
= 0.774.
\end{align}

Our idea is to exploit the relation given by Eq.~\eqref{eq:TKandBbar}
to efficiently and reliably evaluate the Kondo temperature
from a data set of magnetic field dependence of the measured conductance
in the Kondo regime.
We apply Eq.~\eqref{eq:TKandBbar} to an experimental data set
in Sec. \ref{subsec:Application}
and demonstrate evaluation of the Kondo temperature.
We note that the magnetization at $B = \bar{B}$
also takes half of the value at the fully polarized limit
$\left. \langle m_{\rm d}^{} \rangle \right|_{B\to\infty}^{} =1$,
as
\begin{align}
\left. \langle m_{\rm d}^{} \rangle \right|_{B=\bar{B}}^{}
= \frac{1}{2}
\end{align}
for large $U$ as seen in Fig.~\ref{fig:G-B_log}(b).
This relation of the magnetization is useful to evaluate the Kondo temperature of systems in which conductance is
challenging
to be observed,
for example, the Kondo effect in cold atoms
\cite{PhysRevLett.111.135301,PhysRevA.99.032707,PhysRevLett.123.143002}.

\subsection{Empirical formula for magnetic field dependent conductance}

We also present an empirical formula
for the magnetic field dependence of the linear conductance
similar to the temperature dependence given in Eq.~\eqref{eqn:empT}.
We introduce
a formula of magnetic-field dependence of the linear conductance
that takes a similar form to the temperature dependence given in Eq.~\eqref{eqn:empT}, as
\begin{align}
    G_{\rm emp}^B (B) =
    G_0^{} \left[
    1+\left( \frac{B}{\widetilde{B}_{}^{\rm emp}} \right)^2
    \right]^{- t}
    \, .
    \label{eqn:empB}
\end{align}
Because we set that the empirical formula takes half of the zero field value
at $B=\bar{B}$ as $G_{\rm emp}^{B} (\bar{B} ) = G_0^{}/2$,
the Kondo temperature relates to
$\widetilde{B}_{}^{\rm emp}= \bar{B}/ \sqrt{2^{1/t}-1}$
via Eq. \eqref{eq:TKandBbar}.
We numerically determine the value of the fitting parameter $t$ as
\begin{align}
    t = 0.498
    \, ,
\end{align}
using
the least squares
fitting
to the universal part
($\mu_{\rm B}^{} B < T_{\rm K}^{}$) of the conductance curve
of the spin-$\frac{1}{2}$ Kondo model:
We here prepared 2000 data samples of the conductance for equally spaced magnetic fields
from $\mu_{\rm B}^{} B=0$ to $T_{\rm K}^{}$,
using the exact solution of the spin-$\frac{1}{2}$ Kondo model.
We note that, the parameter $t$ is independent of the gate-voltage
as long as the system is in the Kondo regime or the Kondo plateau
in which the charge fluctuation is suppressed
$\langle n_{\rm d}^{} \rangle \simeq 1.0$.

\begin{figure}[tb]
\includegraphics[scale =0.45]{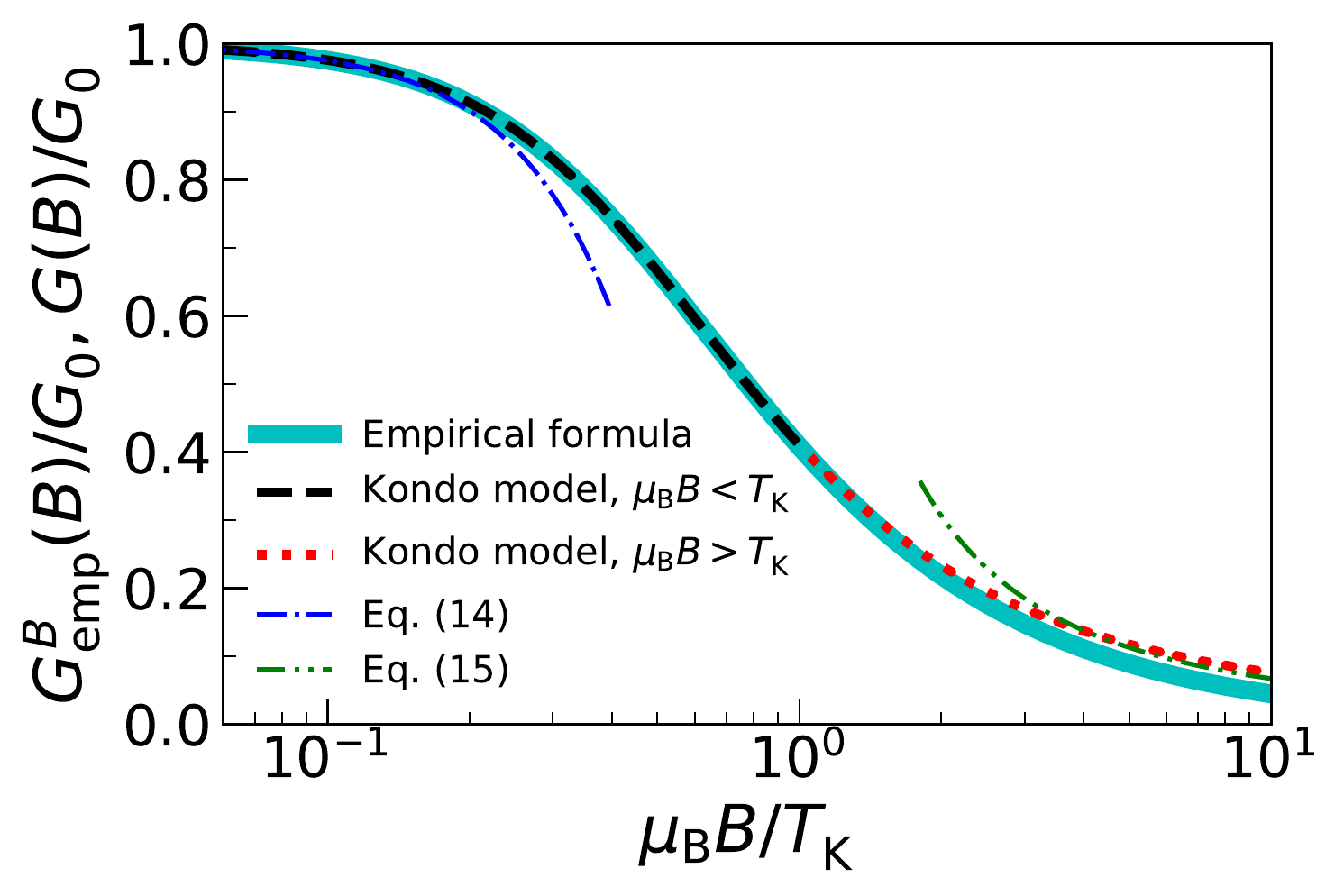}
\caption{ \label{fig:G-B_emp}
The linear conductance calculated by the empirical formula
for the magnetic field dependence,
given in Eq.~\eqref{eqn:empB} (the solid line),
and comparison with that of the exact solution of the Kondo model
(dashed line for $\mu_{\rm B}^{}B < T_K$
and dotted line for $\mu_{\rm B}^{}B > T_K$).
The parameter $t$ of the empirical formula is determined
to fit the universal part ($\mu_{\rm B}^{}B<T_{\rm K}^{}$) given by the exact solution.
The dashed-dotted line and the dashed double-dotted line are asymptotes given by
Eq.~\eqref{eqn:asympototicG4LF} for
$\mu_{\rm B}^{} B\ll T_{\rm K}^{}$
and Eq.~\eqref{eqn:asympototicG4HF} for
$\mu_{\rm B}^{} B\gg T_{\rm K}^{}$,
respectively.
}
\end{figure}

We compare the obtained empirical conductance
with the exact solution of the Kondo model
in Fig.~\ref{fig:G-B_emp}.
Indeed, the empirical equation and the exact solution agree well at magnetic fields below the Kondo temperature
$\mu_{\rm B}^{}B < T_{\rm K}^{} $
while two curves gradually deviate from each other at the higher fields
$\mu_{\rm B}^{}B \gtrsim 2 T_{\rm K}^{} $.

The empirical formula asymptotically behaves at low fields
$\mu_{\rm B}^{} B \ll T_{\rm K}^{}$ as
\begin{align}
& G_{\rm emp}^{B}(B)
\nonumber \\
& =
G_0^{} \left[
1 - \frac{t (2^{1/t} - 1 )}{c_B^2} \left( \frac{\mu_{\rm B}^{} B}{T_{\rm K}^{} } \right)_{}^2 \right]
+ {\cal O} \left( \left( \frac{\mu_{\rm B}^{} B}{T_{\rm K}^{} } \right)_{}^4 \right)
\nonumber \\
& = G_0^{} \left[
1 - 2.51 \times \left( \frac{\mu_{\rm B}^{} B}{T_{\rm K}^{} } \right)_{}^2 \right]
+ {\cal O} \left( \left( \frac{\mu_{\rm B}^{} B}{T_{\rm K}^{} } \right)_{}^4 \right).
\end{align}
This asymptotic form of the empirical formula agrees reasonably with the exact one given in Eq.~\eqref{eqn:asympototicG4LF}
since $\pi^2/4 = 2.467\cdots$. 
At high magnetic fields
$\mu_{\rm B}^{} B \gg T_{\rm K}^{}$,
the empirical formula asymptotically behaves as
\begin{align}
G_{\rm emp}^{B}(B) & \sim G_0^{}
\left( \frac{2^{1/t}-1}{c_B^2} \right)^{-t}
\left( \frac{\mu_{\rm B}^{} B}{T_{\rm K}^{} } \right)_{}^{-2t}
\nonumber \\
& = 0.447 \times G_0^{} \left( \frac{\mu_{\rm B}^{} B}{T_{\rm K}^{} } \right)_{}^{-1.02}
\, .
\label{eqn:EFatHF}
\end{align}
This is not consistent with the exact asymptotic form
given in Eq.~\eqref{eqn:asympototicG4HF}.
This tendency is clearly seen in Fig.~\ref{fig:G-B_emp} at magnetic fields of $\mu_{\rm B}^{} B \sim 10 T_{\rm K}^{}$,
where the deviation of the empirical formula from the exact one gradually develops.
Nevertheless, the empirical formula has a practical merit
as it is a simple function and agrees reasonably well
at $\mu_{\rm B}^{} B \lesssim 2 T_{\rm K}^{}$ with the exact result
for which some extra calculations are needed to obtain the conductance
from Eqs.~\eqref{eqn:ConductanceInMFAtPHS}-\eqref{eqn:md4largeB}.

We exploit the magnetic field dependence of the ground state
to estimate the Kondo temperature
while Goldhaber-Gordon uses the temperature dependence of the excited states.
In the next section, we apply our method given in Eq.~\eqref{eq:TKandBbar}
and Goldhaber-Gordon's empirical formula to experimental data to evaluate the Kondo temperatures.
Then, we compare the results obtained by the two different procedures.
In our experiments,
we estimate the interaction strength to be
$2 \lesssim \frac{U}{\pi\Delta} \lesssim 3$
from observation of the Coulomb diamonds.
Correspondingly, we adopt $\eta(U)=1.1$
following the one estimated in the experiment by Kretinin {\it et al.}
\cite{PhysRevB.84.245316}.
Then, the Kondo temperatures estimated by the two methods
are consistent in our experiments.

\subsection{Application to experimental data}\label{subsec:Application}

\begin{figure*}[tb]
\includegraphics[scale =0.6]{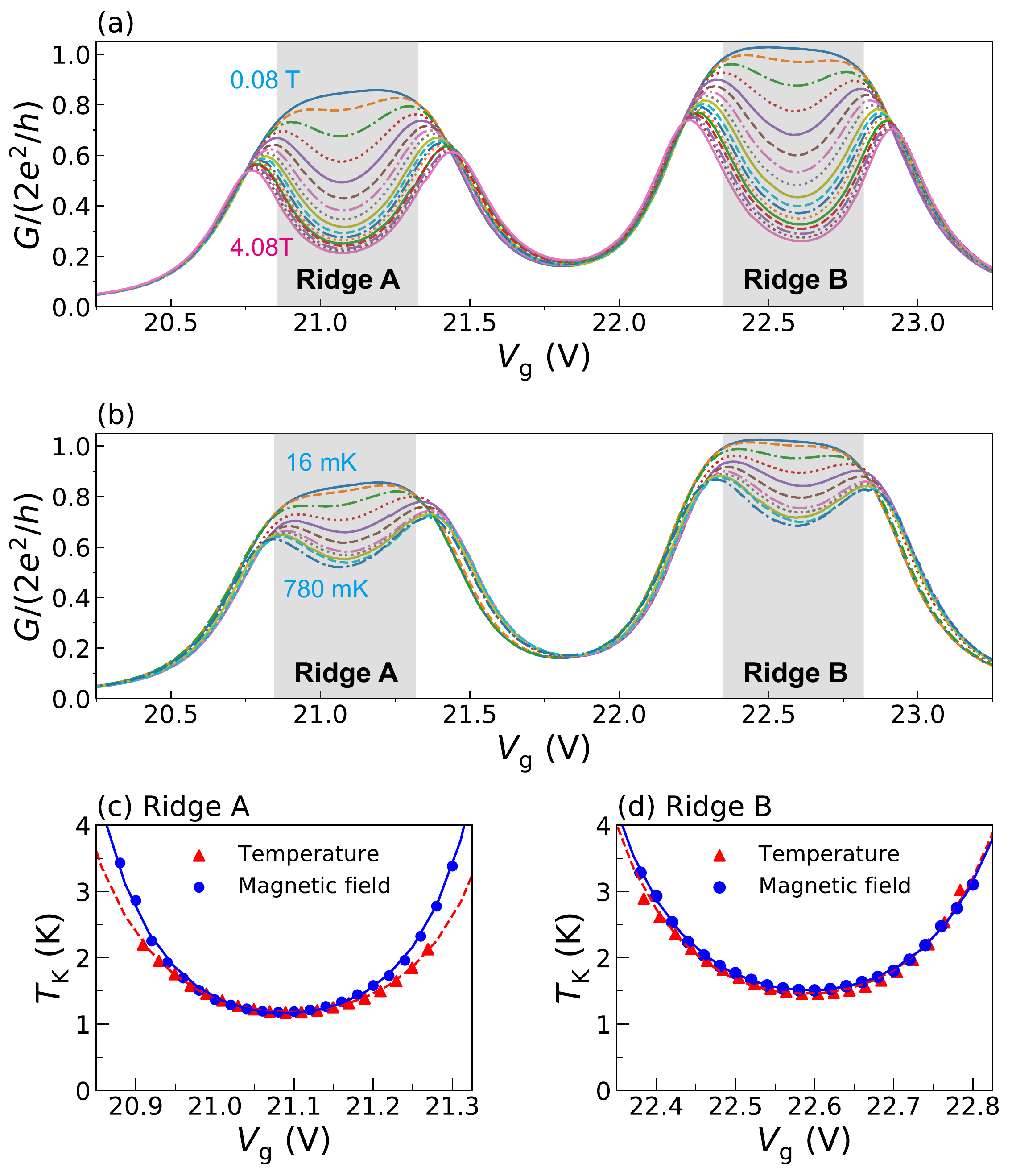}
\caption{\label{fig:exp_G_TK-Vg} Experimentally measured linear conductance $G$
as a function of gate voltage $V_{\rm g}^{}$
(a) at 16~mK
with magnetic fields from 0.08~T to 4.08~T in 0.25~T steps, and
(b) at 0.08~T and
for several temperatures
(16~mK, 80~mK, 180~mK, 280~mK, 380~mK, 480~mK, 580~mK, 630~mK, 680~mK, 730~mK, and 780~mK).
The left conductance ridge around $V_{\rm g}^{}\sim 21.1$~V
and right one around $V_{\rm g}^{}\sim 22.6$~V are labeled
as Ridge A and B, respectively.
(c and d) Evaluated Kondo temperature $T_{\rm K}^{}$ for Ridge A and B, respectively.
The Kondo temperature is evaluated
by using Eq.~\eqref{eq:TKandBbar} for the magnetic field dependence of the conductance ($\bullet$)
and the empirical formula given in Eq.~\eqref{eqn:empT} ($\blacktriangle$) with
$\eta(U)=1.1$ for the temperature dependence of the conductance.
The solid and dashed lines are the fitting curves
given by Eq.~\eqref{eqn:FittingCurve}. }
\end{figure*}

We demonstrate an application of Eq.~\eqref{eq:TKandBbar}
to experimental data of conductance measured in a carbon nanotube quantum dot connected
to two aluminum leads
in a dilution refrigerator
\cite{Nat.Phys.12.230, PhysRevLett.118.196803,PhysRevLett.121.247703, Nat.Comm.12.3233}
to evaluate the Kondo temperature of the dot.
We show the linear conductance as a function of the gate voltage
corresponding to the dot level
with varying applied magnetic field from 0.08~T to 4.08~T at 16~mK
in Fig.~\ref{fig:exp_G_TK-Vg}(a),
and for temperatures from 16~mK to 780~mK at 0.08~T
in Fig.~\ref{fig:exp_G_TK-Vg}(b).
A small finite magnetic field remains even at the minimum magnetic field
to kill the superconductivity in the aluminum leads.
There are two regions in which the spin Kondo effect enhances the conductance
around $V_{\rm g}^{} \sim$ 21.1~V and $V_{\rm g}^{} \sim$ 22.6~V in 
Figs.~\ref{fig:exp_G_TK-Vg}(a) and \ref{fig:exp_G_TK-Vg}(b).
We refer to these regions as Ridge A and Ridge B
for the convenience of discussion.
The value of the gate voltage for each ridge
corresponds to the dot level through
$\epsilon_{\rm d}^{} = \alpha \, U \left( V_{\rm g}^{} - V_{0}^{}\right)$
with the $\alpha$-factor
that presents the lever arm
to tune the dot level.

As seen in Fig.~\ref{fig:exp_G_TK-Vg}(a),
an increase in the magnetic field suppresses the Kondo effect
and it results in decreases in the conductance
in both Ridges A
and B.
Because of the shapes of Ridges A and B,
it is found that
the Kondo effect fully grows the conductance at the lowest magnetic field
in the experiment.
Thus the Kondo effect in both ridges developed enough even at a nonzero small field
$B= 0.08~{\rm T}$,
and we consider that the conductance reaches the value of the low-temperature limit
$G_0^{}$.
The height of the ridges also turns out that
the lead-dot connection is almost symmetric
$\Gamma_L^{} \simeq \Gamma_R^{}$
in Ridge B
while it is asymmetric
$\Gamma_L^{} \neq \Gamma_R^{}$
in Ridge A.

Here, let us evaluate the Kondo temperature of the two ridges
to the experimental data of the magnetic field dependence of the linear conductance shown in Fig.~\ref{fig:exp_G_TK-Vg}(a).
Specifically, we estimated $\bar{B}$
by applying linear interpolation to the two
magnetic-field $B$ points that give the conductance
$G(B) \simeq G(B_0^{} ={\rm 0.08~T})/2$
to determine the Kondo temperature
through Eq.~\eqref{eq:TKandBbar}.
We also carried out this procedure for several gate voltages in the Kondo regime,
for both Ridges A and B shown in Fig.~\ref{fig:exp_G_TK-Vg}(a).
We take the $g$-factor of our carbon nanotube quantum dot as $g=2$
similarly to this theoretical analysis so far
\cite{RevModPhys.87.703}.
We plot the obtained values of the Kondo temperature $T_{\rm K}^{}$
as functions of the gate voltage $V_{\rm g}^{}$
in Ridges A and B
in Figs.~\ref{fig:exp_G_TK-Vg}(c) and \ref{fig:exp_G_TK-Vg}(d), respectively.
For comparison,
we also evaluate the Kondo temperature
from the temperature dependence of the conductance shown
in Fig.~\ref{fig:exp_G_TK-Vg}(b)
by using the empirical formula given in Eq.~\eqref{eqn:empT}.
The results are also plotted
in Figs.~\ref{fig:exp_G_TK-Vg}(c) and \ref{fig:exp_G_TK-Vg}(d).
We here use Eq.~\eqref{eqn:modeldependence} and take $\eta(U) = 1.1$
in both Ridges A and B to compare the empirical Kondo temperature $T_{\rm K}^{\rm emp}$ to
the Kondo temperature evaluated from the ground state $T_{\rm K}^{}$,
similarly to Kretinin {\it et al.}
\cite{PhysRevB.84.245316}.
As seen in Figs.~\ref{fig:exp_G_TK-Vg}(c) and \ref{fig:exp_G_TK-Vg}(d),
the Kondo temperatures that are evaluated in the two different ways
agree very well in both ridges.

\begin{figure*}[tb]
\includegraphics[scale =0.7]{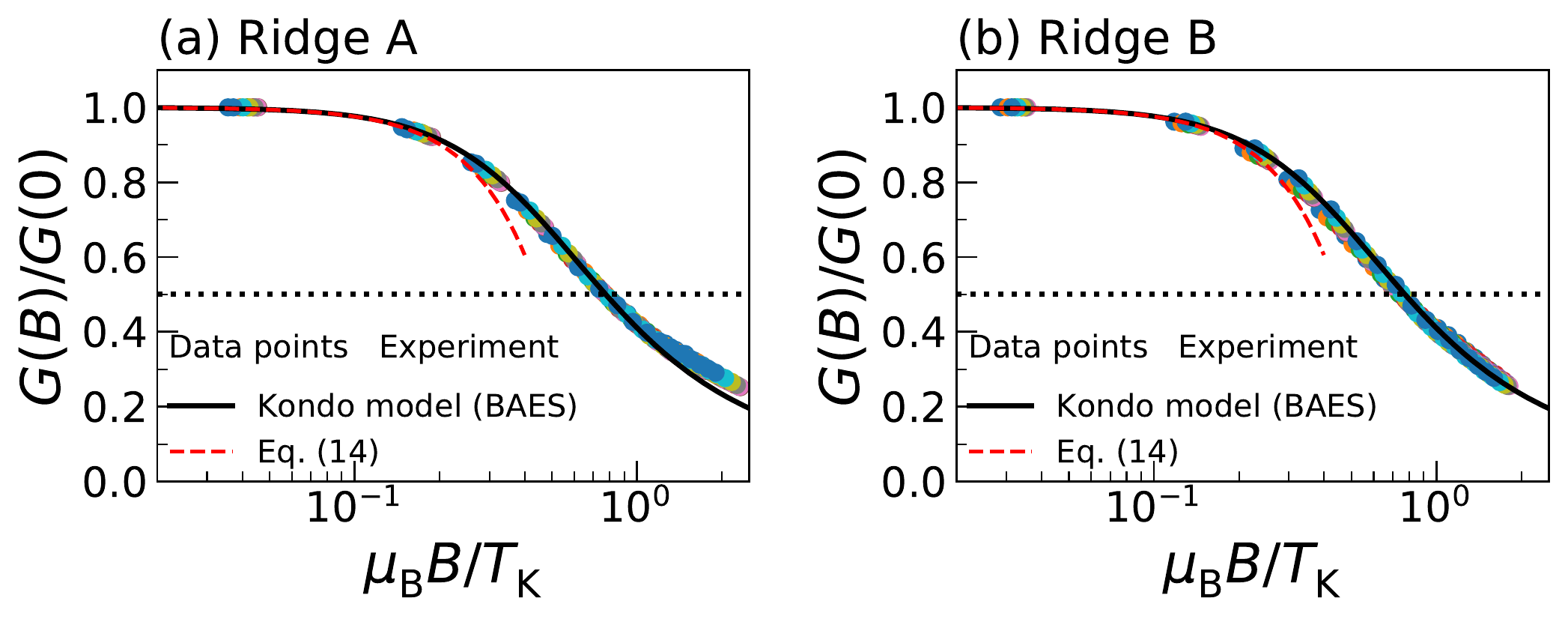}
\caption{\label{fig:exp_G-B} The linear conductance
as a function of the applied magnetic field scaled by the Kondo temperature
for several choices of gate voltage $V_{\rm g}^{}$ corresponding to the Kondo regime.
In (a) Ridge A with an asymmetric lead-dot connection
and (b) Ridge B with the symmetric lead-dot connection.
The magnetic field is scaled by the Kondo temperature
given in Fig.~\ref{fig:exp_G_TK-Vg}.
The solid line is the conductance calculated by the Bethe ansatz exact solution for the Kondo model,
which is normalized by the value at zero fields $B=0$,
$G(0)=G_0^{}$.
The data points are the experimentally observed conductance
(a) for $V_{\rm g}^{}$ from 20.98~V to 21.18~V in 0.02~V steps in Ridge A
and (b) for $V_{\rm g}^{}$ from 22.48~V to 22.68~V in 0.02~V steps
in Ridge B.
The experimental conductance is normalized
by its value at the lowest field, 0.08T, instead of its zero field value.
The values of the Kondo temperature, determined by the magnetic field dependence for each gate voltage, are used.
The dashed lines are the formula for the low fields,
given by Eq.~\eqref{eqn:asympototicG4LF}.
}
\end{figure*}

Finally,
we also examine the universal scaling property
in magnetic field dependence of the linear conductance
to test the validity of the evaluation of the Kondo temperature
in our procedure.
The linear conductance as a function of the magnetic field scaled
by the Kondo temperature in Ridges A and B are plotted
in Figs.~\ref{fig:exp_G-B}(a) and \ref{fig:exp_G-B}(b), respectively.
In the figure,
we scale the magnetic field by the Kondo temperature evaluated
by Eq.~\eqref{eq:TKandBbar} for each $V_{\rm g}^{}$.
We find that,
for smaller magnetic fields than the Kondo temperature $\mu_{\rm B}^{} B \lesssim T_{\rm K}^{}$,
the universal scaling property works very well in both ridges.
This universal scaling is observed not only
at the center of the ridge
but also on a broader range of $V_{\rm g}^{}$
from 20.98~V to 21.18~V for Ridge A, and from 22.48~V to 22.68~V for Ridge B,
where
the charge fluctuation is suppressed
as $\langle n_{\rm d}^{} \rangle \simeq 1.0$
and $\chi_{\rm c}^{}/\chi_{\rm s}^{} \ll 1$.
We here conclude that 
our procedure to evaluate the Kondo temperature from the magnetic field
dependence works very well.
We note that outside of the gate voltage range
shown in Figs.~\ref{fig:exp_G-B}(a) and \ref{fig:exp_G-B}(b),
the charge fluctuation becomes larger and the scaling universality is gradually broken.
We also reported that the universal $T/T_{\rm K}^{}$ scaling of the conductance
works very well in the Kondo plateau of Ridge B,
in the supplemental information of Ref.~\onlinecite{Nat.Phys.12.230}.

We examined universal scaling on magnetic field dependence of the nonlinear conductance
in Ridge B with the symmetric lead-dot connection
$\Gamma_L^{} = \Gamma_R^{}$
in our previous paper
\cite{Nat.Comm.12.3233}.
The universal scaling curve of the nonlinear conductance is given in
an intricate form of the lead-dot connections $\Gamma_{\alpha}^{}$ 
\cite{PhysRevB.104.235147}.
However,
the linear conductance depends on the asymmetry of
the lead-dot connections only through
the allover coefficient $G_0^{}$
that the zero field value
determines as $G_0^{} = G(0)$.
This enables one to eliminate the asymmetry of
the lead-dot connections
from the universal curve of the linear conductance,
as seen in the discussion so far.
Thus, the universal curve of the linear conductance is less demanding,
and our procedure is suited to evaluating the Kondo temperature.

\section{summary}\label{sec:summary}

We have investigated a universal scaling property of the bias-voltage linear response
to the applied magnetic field
in a quantum dot,
using the exact solution for the spin-$\frac{1}{2}$ Kondo model
and the numerical renormalization group approach for the spin-$\frac{1}{2}$ Anderson impurity model.
Then, we showed a procedure
to estimate the Kondo temperature for experimental data sets
using the universal curve of the magnetic field dependence of the linear conductance
derived by the Bethe ansatz exact solution of the spin-$\frac{1}{2}$ Kondo model.
We also demonstrate the application of our procedure to experimental data of linear conductance
through a carbon nanotube quantum dot connected to two aluminum leads.

An advantage of our procedure to evaluate the Kondo temperature
is that it is applicable to systems in which the temperature is
difficult to control,
such as ultracold atoms.
Although conductance is challenging
to observe in cold atom systems,
the universal magnetic field dependence of the magnetization can be used to estimate the Kondo temperature as in the linear conductance.
Furthermore,
the magnetic field can be varied on a broader energy range with keeping the system stable
compared with the system temperature in semiconductor systems.
Temperature can also be tuned in a broader energy range in principle,
but the thermal cycle with a large temperature range usually
gives rise to some changes on system parameters.
As seen in our experiment,
the applied magnetic field is often varied with a broader energy range
than the temperature.

In the paper, the $g$-factor has been fixed at $g=2$.
However, in some actual systems, the $g$-factor is unknown.
The $g$-factor can be determined by estimating the Kondo temperature
in addition to the magnetic field dependence in other ways.
For example,
we can estimate $\bar{B}$ from magnetic field dependence of conductance
and the Kondo temperature from temperature dependence.
Then we can estimate the $g$-factor
by using Eq.~\eqref{eq:TKandBbar}.

We can also extend this method to determine
the Kondo temperature of systems with orbital degeneracy.
However,
we need to carefully introduce the arrangement of orbitals into the model,
because the response of orbital to applied magnetic fields depends
much on materials.

\begin{acknowledgments}
This work was partially supported by
JSPS KAKENHI Grants No. JP18K03495, No. JP18J10205, No. JP19H00656, No. JP19H05826, No. JP19K14630, No. JP20K03807, No. JP21K03415, No. JP22H01964, No. JP26220711, and No. JP23K03284,
JST CREST Grant No. JPMJCR1876,
and the French program ANR JETS (Grant No. ANR-16-CE30-0029-01).
KM was supported by JST Establishment of University Fellowships towards the Creation of Science Technology Innovation Grant No. JPMJFS2138.
\end{acknowledgments}

\appendix

\section{Kondo model}\label{Appendix:KondoModel}
We decompose the conduction electrons in left and right leads
in the Anderson impurity model \eqref{eqn:AIM}
to the connected and disconnected electrons to the dot as
\begin{align}
    c_{\varepsilon\sigma}^{} = &
    \frac{v_L^{}}{v} c_{\varepsilon L \sigma}^{}
    + \frac{v_R^{}}{v} c_{\varepsilon R \sigma}^{}
    \, , \\
    \bar{c}_{\varepsilon\sigma}^{} = &
    - \frac{v_R^{*}}{v} c_{\varepsilon L \sigma}^{}
    + \frac{v_L^{*}}{v} c_{\varepsilon R \sigma}^{}
\end{align}
with
$v = \sqrt{|v_L^{}|^2 + |v_R^{}|^2}$.
This unitary transformation rewrite the electron tunneling and conduction electron parts of the Hamiltonian as
\begin{align}
    \cal{H}_{\rm c}^{} = &
    \sum_{\sigma} \int_{-D}^{D} d\varepsilon \, \varepsilon
    (c_{\varepsilon\sigma}^{\dagger} c_{\varepsilon\sigma}^{}
    + \bar{c}_{\varepsilon\sigma}^{\dagger} \bar{c}_{\varepsilon\sigma}^{})
    \, , \\
    \cal{H}_{\rm T}^{} = &
    	\sum_{\sigma} v \left(
	d_{\sigma}^{\dagger} \psi_{\sigma}
	+ \psi_{ \sigma}^{\dagger} d_{\sigma}^{}
	\right)
\end{align}
with
$\psi_{\sigma}^{} = \int_{-D}^{D} d\varepsilon \sqrt{\rho_{\rm c}^{}} c_{\varepsilon\sigma}^{}$.
Thus the disconnected state does not contribute to the dot state and is ignored.
By applying the Schriefer-Wolf transformation to the Anderson impurity model with
$U\gg \Delta$ and $\langle n_{\rm d}^{} \rangle \sim 1$
\cite{PhysRev.149.491},
we get the Kondo model 
\begin{align}
    {\cal H}_{\rm K}^{} = & \sum_{\sigma} \int_{-D}^{D} d\varepsilon \, \varepsilon \,
    c_{\varepsilon\sigma}^{\dagger} c_{\varepsilon\sigma}^{}
    \nonumber \\
    & + J_{\rm K}\sum_{\sigma\sigma'} \frac{1}{2}
    \psi_{\sigma}^{\dagger} {\bm \sigma}_{\sigma\sigma'}^{} \psi_{\sigma'}^{} \cdot {\bm S}_{\rm d}
    -\frac{1}{2}g\mu_{\rm B}^{} B S_{\rm d}^{z}
\end{align}
with an antiferromagnetic exchange interaction
\begin{align}
    J_{\rm K}^{} = -2v^2 \frac{U}{\epsilon_{\rm d}^{}(\epsilon_{\rm d}^{}+U)}
    > 0 \, ,
\end{align}
Here, ${\bm \sigma}_{\sigma\sigma'}^{}$ is the Pauli spin matrix,
${\bm S}_{\rm d}^{}$ is the spin operator at the dot site, and
$S_{\rm d}^{z}$ is the $z$-component of ${\bm S}_{\rm d}^{}$.

We calculate the magnetization at the dot site
$\langle m_{\rm d}^{} \rangle$
using the Bethe ansatz exact solution of this Kondo model in this paper.

\section{Fitting function of Figs.~\ref{fig:exp_G_TK-Vg}(c) and \ref{fig:exp_G_TK-Vg}(d)}
We use an exponential function with a quadratic as the exponent,
\begin{align}
    F_{\rm fitting}^{} (V_{\rm g}^{}) =
    \exp \left[ c_2^{} V_{\rm g}^2 + c_1^{} V_{\rm g}^{} + c_0^{} \right]
    \label{eqn:FittingCurve}
\end{align}
as the fitting curve for the gate-voltage dependence of the Kondo temperature.
Here, $c_0^{}, c_1^{}$, and $c_2^{}$ are the fitting parameters.
This function is chosen to be consistent with the asymptotic form of the gate-voltage dependence
of the Kondo temperature derived by the scaling analysis
\cite{PhysRevLett.40.416}.
As seen
in Figs.~\ref{fig:exp_G_TK-Vg}(c) and \ref{fig:exp_G_TK-Vg}(d),
the curve given by Eq.~\eqref{eqn:FittingCurve} fits the evaluated Kondo temperatures very well.
Therefore, the gate voltage dependence of the evaluated Kondo temperature via Eq.~\eqref{eq:TKandBbar}
agrees with the asymptotic behavior that is theoretically expected.
We note that, in principle,
$U$ and $\Delta$ can be deduced from the fitting parameters
for systems with large Coulomb interaction $\frac{U}{\pi\Delta} \gtrsim 3$.
However, the error of the fitting parameters becomes large
because the Kondo temperature is given in the exponential form and deduced interaction is small,
$2 \lesssim \frac{U}{\pi\Delta} \lesssim 3$.



%

\end{document}